\documentclass[10pt, conference, letterpaper]{IEEEtran}
\usepackage{amsmath, amsfonts,  amssymb, color, textcomp}
\usepackage[all]{xy}
\usepackage{graphicx}
\usepackage{latexsym}
\usepackage{epsfig}
\usepackage{bm}
\usepackage{xspace}
\usepackage{url}
 \usepackage[footnotesize]{caption}

\newcommand{\btb}{\mathbf{B}_{tb}^{(\lambda)}}
\newcommand{\bterm}{\mathbf{B}_{[0,L-1]}}
\newcommand{\binf}{\mathbf{B}_{[0,\infty]}}

\long\def\symbolfootnote[#1]#2{\begingroup%
\def\thefootnote{\fnsymbol{footnote}}\footnote[#1]{#2}\endgroup}

\renewcommand{\baselinestretch}{0.96}
\begin{document}

\title{On the Minimum Distance of Generalized \\Spatially Coupled LDPC Codes\vspace{0mm}}
\author{
\authorblockN{David G. M. Mitchell$^*$, Michael Lentmaier$^\dag$, and Daniel J. Costello, Jr.$^*$}
\authorblockA{$^*$Dept. of Electrical Engineering, University of Notre Dame, Notre Dame,
Indiana, USA,\\
\{david.mitchell, costello.2\}@nd.edu\\
$^\dag$Dept. of Electrical and Information Technology, Lund University, Lund, Sweden\\
Michael.Lentmaier@eit.lth.se}\vspace{-6mm}}

\maketitle
\begin{abstract}
Families of generalized spatially-coupled low-density parity-check (GSC-LDPC) code ensembles can be formed by terminating protograph-based generalized LDPC convolutional (GLDPCC) codes. It has previously been shown that ensembles of GSC-LDPC codes constructed from a protograph have better iterative decoding thresholds than their block code counterparts, and that, for large termination lengths, their thresholds coincide with the maximum a-posteriori (MAP) decoding threshold of the underlying generalized LDPC block code ensemble. Here we show that, in addition to their excellent iterative decoding thresholds, ensembles of GSC-LDPC codes are asymptotically good and have large minimum distance growth rates.

\end{abstract}

\section{Introduction}
%\symbolfootnote[0]{This work was partially supported by NSF Grant CCF-1161754.}
Low-density parity-check convolutional (LDPCC) codes \cite{fz99} have been shown to be capable of achieving capacity-approaching performance with iterative message-passing decoding \cite{psvc11}. The excellent iterative decoding thresholds \cite{lscz10,lfzc09} that these codes display has been attributed to the \emph{threshold saturation} effect \cite{kru11,kru12}. In addition to good threshold performance, it can also be shown that the minimum free distance typical of most members of these LDPCC code ensembles grows linearly with the constraint length as the constraint length tends to infinity, i.e., they are {\em asymptotically good} \cite{stl+07,mpc13}. A large free distance growth rate indicates that codes randomly drawn from the ensemble should have a low error floor under maximum likelihood (ML) decoding. 

Generalized LDPC (GLDPC) block codes were first proposed by Tanner \cite{tan81} and have been shown to possess many desirable features, such as large minimum distance \cite{lz99,bpz99} and good iterative decoding performance \cite{lrc08}. Following this construction, more complicated constraints than a single parity-check (SPC) constraint are permitted. In other words, a constraint node with $n$ inputs can represent an arbitrary $(n,k)$ linear block code.   In \cite{lf10}, Lentmaier and Fettweis showed that ensembles of generalized terminated LDPCC codes, called generalized spatially-coupled LDPC codes (GSC-LDPC) codes, constructed from a protograph have better iterative decoding thresholds than their block code counterparts, and that, for large termination lengths, their thresholds coincide with the maximum a-posteriori (MAP) decoding threshold of the underlying GLDPC block code ensemble.

In this paper, using weight enumerator evaluation techniques presented by Abu-Surra, Divsalar, and Ryan \cite{adr11}, we study the asymptotic weight spectrum of GSC-LDPC code ensembles. We show, using a $(2,7)$-regular GLDPC block code with $(7,4)$ Hamming code constraints as an example, that the corresponding GSC-LDPC code ensembles are asymptotically good and have large minimum distance growth rates. As the termination length increases, we obtain a family of codes with capacity approaching iterative decoding thresholds and declining minimum distance growth rates. However, since these are convolutional codes, a more appropriate distance measure for assessing the ML  decoding performance of such code ensembles is the \emph{free distance} growth rate of the associated ensemble of periodically time-varying generalized LDPC convolutional (GLDPCC) codes. Consequently, in the final part of the paper, we show that the terminated GSC-LDPC code ensembles can be used to obtain an upper bound on the free distance growth rate of ensembles of periodically time-varying GLDPCC codes. The free distance growth rate can also be bounded below by using ensembles of tail-biting GLDPCC codes using a similar technique to one previously presented for LDPCC code ensembles with SPC constraints \cite{tzc10,mpc13}. By comparing and evaluating these bounds we find that, for a sufficiently large period, the bounds coincide, giving us exact values for the GLDPCC code ensemble free distance growth rates. \vspace{-2mm}

\section{Background}
\begin{figure}[h]
\begin{center}
\includegraphics[width=2.4in]{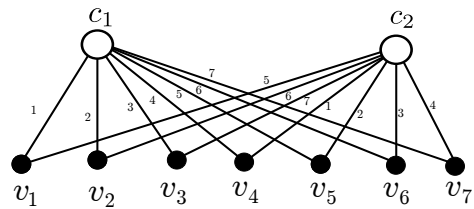}
\end{center}
\caption{Protograph of a $(2,7)$-regular GLDPC block code. The white circles represent generalized constraint nodes and the black circles represent variable nodes. The labels on the edges indicate the corresponding columns of the parity check matrix of the generalized constraint code.}\label{fig:27ham}\vspace{-3mm}
\end{figure}

A protograph \cite{tho03} is a small bipartite graph that connects a set of $n_v$ variable nodes $V=\{v_1,\ldots,v_{n_v}\}$ to a set of $n_c$ generalized constraint nodes $C=\{c_1,\ldots,c_{n_c}\}$ by a set of edges $E$. In a protograph-based GLDPC code, each constraint node $c_m$ can represent an arbitrary block code of length $n_{c_m}$. Figure \ref{fig:27ham} displays the protograph of a  $(2,7)$-regular GLDPC block code.

A protograph can be represented by means of an $n_c\times n_v$ bi-adjacency matrix $\mathbf{B}$, which is called the \emph{base matrix} of the protograph. The entry in row $i$ and column $j$ of $\mathbf{B}$ is equal to the number of edges that connect nodes $c_i$ and $v_j$. The base matrix of the protograph in Fig. \ref{fig:27ham} is given by
\begin{equation}
\mathbf{B}=\left[
\begin{array}{ccccccc}
1 & 1 & 1 & 1 & 1 & 1 & 1\\
1 & 1 & 1 & 1 & 1 & 1 & 1
\end{array}\right].\label{hammingbase}
\end{equation}

In order to construct ensembles of  protograph-based GLDPC codes, a protograph can be interpreted as a template for the Tanner graph of a derived code, which can be obtained by a copy-and-permute operation \cite{tho03}. The protograph is lifted by replicating each node $N$ times and the edges are permuted among these replicated nodes in such a way that the structure of the original graph is preserved. Allowing the permutations to vary over all $N!$ possible choices results in an ensemble of GLDPC block codes.

\subsection{Convolutional protographs}\label{sec:convproto}
An unterminated GLDPCC code can be described by a {\em convolutional protograph} \cite{lfzc09} with base matrix
\begin{equation}\label{convbase}\mathbf{B}_{[0,\infty]}=\left[
\begin{array}{cccccc}
\mathbf{B}_{0} &   & \\
\mathbf{B}_{1} &\mathbf{B}_{0} &  \vspace{-2.2mm}\\
\vdots &\mathbf{B}_{1} & \vspace{-2mm}\ddots \\
\mathbf{B}_{m_s} & \vdots& \ddots\\
 & \mathbf{B}_{m_s}& \vspace{-2mm}\\
 & & \ddots\\
\end{array}\right],\end{equation}
where $m_s$ denotes the syndrome former memory of the convolutional code and the $b_c \times b_v$ {\em component base matrices} $\mathbf{B}_{i}$, $i=0,1,\dots,m_s$, represent the edge connections from the $b_v$ variable nodes at time $t$ to the $b_c$ (generalized) constraint nodes at time $t+i$. An ensemble of (in general) time-varying GLDPCC codes can then be formed from $\mathbf{B}_{[0,\infty]}$ using the protograph construction method  described above. The \emph{decoding constraint length} of the resulting ensemble is given as $\nu_s = (m_s + 1)Nb_v$.

Starting from the base matrix $\mathbf{B}$ of a block code ensemble, one can construct GLDPCC code ensembles with the same computation trees. This is achieved by an {\em edge spreading} procedure (see \cite{lfzc09} for details) that divides the edges from each variable node in the base matrix $\mathbf{B}$ among $m_s+1$ component base matrices $\mathbf{B}_i$, $i=0,1,\dots,m_s$, such that the condition $\mathbf{B}_0+\mathbf{B}_1+\cdots+\mathbf{B}_{m_s}=\mathbf{B}$ is satisfied. For example, we could apply the edge spreading technique to the $(2,7)$-regular block base matrix in (\ref{hammingbase}) to obtain the following component base matrices
\begin{eqnarray}
&\mathbf{B_0}=\left[
\begin{array}{ccccccc}
0 & 0 & 0 & 0 & 1 & 1 & 1\\
1 & 1 & 1 & 0 & 0 & 0 & 0
\end{array}\right],\label{b1}\\
&\mathbf{B_1}=\left[
\begin{array}{ccccccc}
1 & 1 & 1 & 1 & 0 & 0 & 0\\
0 & 0 & 0 & 1 & 1 & 1 & 1
\end{array}\right].\label{b2}
\end{eqnarray}
\begin{figure*}[t]
\begin{center}
\includegraphics[width=6.5in]{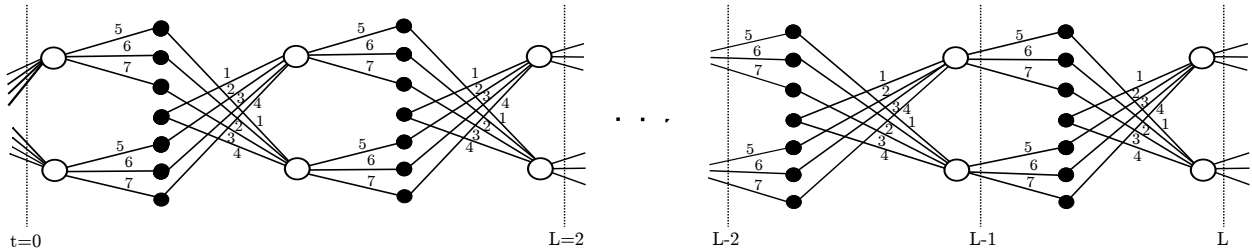}
\end{center}
\caption{Protograph of a $(2,7)$-regular GLDPCC code ensemble. The white circles represent generalized constraint nodes and the black circles represent variable nodes.}\label{fig:scprot}\vspace{-7mm}
\end{figure*}
 
From a convolutional protograph with base matrix $\mathbf{B}_{[0,\infty]}$, we can form a periodically time-varying $N$-fold graph cover with period $T$ by choosing, for the $b_c\times b_v$ submatrices $\mathbf{B}_0,\mathbf{B}_1,\ldots,\mathbf{B}_{m_s}$ in the first $T$ columns of $\mathbf{B}_{[0,\infty]}$, a set of $N\times N$ permutation matrices randomly and independently to form $Nb_c \times Nb_v$ submatrices $\mathbf{H}_0(t),\mathbf{H}_1(t+1),\ldots,\mathbf{H}_{m_s}(t+m_s)$, respectively, for $t=0,1,\ldots,T-1$. These submatrices are then repeated periodically (indefinitely) to form a convolutional parity-check matrix  $\mathbf{H}_{[0,\infty]}$ such that $\mathbf{H}_i(t+T)=\mathbf{H}_i(t)$, $\forall i,t$. An ensemble of periodically time-varying GLDPCC codes with period $T$, rate $R=1-NM_\mathcal{C}b_c/Nb_v=1-M_\mathcal{C}b_c/b_v$, and decoding constraint length $\nu_s=N(m_s+1)b_v$ can then be derived by letting the permutation matrices used to form $\mathbf{H}_0(t),\mathbf{H}_1(t+1),\ldots,\mathbf{H}_{m_s}(t+m_s)$, for $t=0,1,\ldots,T-1$, vary over the $N!$ choices of an $N\times N$ permutation matrix.
\section{Termination of GLDPCC codes}\vspace{-1mm}
Suppose that we start the convolutional code with parity-check matrix defined in $(\ref{convbase})$ at time $t=0$ and terminate it after $L$ time instants. The resulting finite-length base matrix is then given by \vspace{-1mm}
\begin{equation}\label{termbase}\mathbf{B}_{[0,L-1]}=\left[
\begin{array}{ccc}
\mathbf{B}_0 & &\\
%\mathbf{B}_1 & \ddots & \mathbf{B}_1\\
\vdots & \ddots &  \\
\mathbf{B}_{m_s} &  & \mathbf{B}_0 \\
& \ddots & \vdots\\
& & \mathbf{B}_{m_s}
\end{array}\right]_{(L+m_s)b_c \times Lb_v}. \vspace{-1mm}\end{equation}
The matrix $\mathbf{B}_{[0,L-1]}$ can be considered as the base matrix of a terminated protograph-based GLDPCC code, or generalized \emph{spatially-coupled} LDPC (GSC-LDPC) code. This terminated protograph is slightly irregular with lower constraint node degrees at the beginning and end. These shortened constraint nodes are now associated with shortened constraint codes in which the symbols of the missing edges are removed. Note that such a code shortening is equivalent to fixing these removed symbols and assigning an infinite reliability to them. The variable node degrees are not affected by termination.

The parity-check matrix $\mathbf{H}$ of the block code derived from $\mathbf{B}_{[0,L-1]}$ by lifting with some factor $N$  has $Nb_v L$ columns and $(L+m_s)Nb_cM_\mathcal{C}$ rows, where $M_\mathcal{C}$ denotes the number of parity-checks of the constraint code.\footnote{We assume here that each generalized constraint node in the block protograph is of the same type and has $M_\mathcal{C}$ parity-checks. This assumption can be relaxed in general.} It follows that the rate of the GSC-LDPC code is equal to
\begin{equation}R_L =1-\frac{(L+m_s)b_cM_\mathcal{C}-\Delta}{Lb_v} \vspace{-1mm}\end{equation}
for some $\Delta\geq 0$ that accounts for a slight rate increase due to the shortened constraint nodes. If $\mathbf{H}$ has full rank, the rate increase parameter is given by $\Delta = 0$. However, shortened constraint codes at the ends of the graph can cause a reduced rank for $\mathbf{H}$, which slightly increases $R_L$. In this case,  $\Delta>0$ and depends on both the particular constraint code chosen and the degree of shortening. As $L\rightarrow\infty$, the rate $R_L$ converges to the rate of the underlying GLDPC block code with base matrix $\mathbf{B}$. 

The generalized convolutional base matrix $\mathbf{B}_{[0,\infty]}$ can also be terminated using \emph{tail-biting} \cite{st79,mw86}, resulting in a generalized tail-biting LDPC (GTB-LDPC) code. Here, for any $\lambda \geq m_s$, the last $b_cm_s$ rows of the terminated parity-check matrix $\mathbf{B}_{[0,\lambda-1]}$ are removed and added to the first $b_cm_s$ rows to form the $\lambda b_c \times \lambda b_v$ tail-biting parity-check matrix $\mathbf{B}_{tb}^{(\lambda)}$ with tail-biting termination factor $\lambda$:\vspace{-1mm} %. The resulting $\lambda b_c \times \lambda b_v$ tail-biting parity-check matrix $\mathbf{B}_{tb}^{(\lambda)}% with termination factor $\lambda$ is shown below: 
\begin{equation}\label{basetb}
\scalebox{0.9}{\mbox{$\left[\begin{array}{cccccccccc}
\mathbf{B}_0 &  &&& &   &  & \mathbf{B}_{m_s}   & \cdots &  \mathbf{B}_1 \\
\mathbf{B}_1 & \mathbf{B}_0  & &&&  &    &  & \ddots& \vdots \\
\vdots & \vdots  &&&&   &  &  &    & \mathbf{B}_{m_s}  \\
\mathbf{B}_{m_s} & \mathbf{B}_{m_s-1}  & &&&  &  &  &    & \\
 & \mathbf{B}_{m_s}  & &&& \ddots  &  &  &  &  \\
 &   & &&&  & \mathbf{B}_{0} &  &  &    \\
 &   &&&& \ddots  & \vdots & \mathbf{B}_{0}  &  &  \\
 &   & &&& \ddots & \mathbf{B}_{m_s-1}  & \vdots & \ddots &    \\
 &  & &&&& \mathbf{B}_{m_s} & \mathbf{B}_{m_s-1} & \cdots &\mathbf{B}_{0}   \\
\end{array}\right].$}}\end{equation}
\noindent  Note that, if $m_s=1$ and $\lambda = 1$, the tail-biting base matrix is simply the original block code base matrix, i.e., $\mathbf{B}_{tb}^{(1)}=\mathbf{B}$. 
Terminating $\mathbf{B}_{[0,\infty]}$ in such a way preserves the design rate of the ensemble, i.e., $R_\lambda =1-\lambda b_c M_\mathcal{C}/\lambda b_v=1-b_cM_\mathcal{C}/b_v=R$, and we see that $\mathbf{B}_{tb}^{(\lambda)}$ has exactly the same degree distribution as the original block code base matrix $\mathbf{B}$.

\section{Minimum distance analysis of protograph-based GSC-LDPC code ensembles}\label{sec:dist}
In \cite{adr11}, Abu-Surra, Divsalar, and Ryan presented a technique to calculate the average weight enumerator and asymptotic spectral shape function for protograph-based GLDPC code ensembles. The spectral shape function can be used to test if an ensemble is \emph{asymptotically good}, i.e., if the minimum distance typical of most members of the ensemble is at least as large as $\delta_{min}n$, where $\delta_{min}$ is the \emph{minimum distance growth rate} of the ensemble and $n$ is the block length.  

Consider the protograph with generalized constraint nodes shown in Fig. \ref{fig:27ham}. If we suppose the constraint nodes to be $(7,4)$ Hamming codes with parity-check matrix 
$$\mathbf{H}_1=\left[\begin{array}{ccccccc}
1 & 0 & 0 & 1 & 1 & 1 & 0\\
0 & 1 & 0 & 1 & 1 & 0 & 1\\
0 & 0 & 1 & 1 & 0 & 1 & 1
\end{array}\right],$$
then the resulting  ensemble has design rate $R=1/7$, is asymptotically good, and has growth rate $\delta_{min}=0.186$ \cite{adr11}.

%\begin{figure}[t]
%\setcounter{figure}{1}
%\begin{center}
%\includegraphics[width=\columnwidth]{block.eps}
%\end{center}
%\caption{Asymptotic codeword weight enumerator for a rate $R=1/7$ GLDPC code ensemble. }\label{fig:block}
%\end{figure}

We will construct the base matrix of a GSC-LDPC code ensemble using (\ref{convbase}) and component base matrices (\ref{b1}) and (\ref{b2}). The resulting protograph is shown in Fig. \ref{fig:scprot}. We use the $(7,4)$ Hamming code with parity-check matrix $\mathbf{H}_1$ for the generalized constraint nodes. The numbers on the edges of the protograph in Fig. \ref{fig:scprot} indicate which columns of $\mathbf{H}_1$ (or shortened version of $\mathbf{H}_1$) the nodes are connected to. After termination, the resulting ensemble corresponds to a GSC-LDPC code ensemble. The design rate of the ensemble is given as
\begin{equation}\label{ratesc}
R_L = 1-\frac{6(L+1)-2}{7L}.
\end{equation}
Note that $\Delta=2$ in this example because the two leftmost (shortened) constraint nodes in Fig. \ref{fig:scprot} correspond to shortened codes with rate $1/3$, i.e., the number of parity-checks in these two constraint nodes is $M_\mathcal{C}=2$, while all of the other constraint nodes have $M_\mathcal{C}=3$ parity-checks. These ensembles were shown to have thresholds numerically indistinguishable from capacity as $L\rightarrow\infty$ in \cite{lf10}.

The evaluation of the asymptotic weight enumerators for GSC-LDPC codes is complex, since the conjecture regarding simplification of the numerical evaluation proposed in  \cite{adr11} cannot immediately be applied to these ensembles. This conjecture relies on grouping together nodes of the same type and optimizing them together. However, in the GSC-LDPC case, nodes from different time instants must be optimized separately, even if they are of the same type.% A modified conjecture that can be used to simplify the weight enumerator evaluation of GSC-LDPC codes is the subject of ongoing work. 

\begin{figure}[h]
\begin{center}
\includegraphics[width=\columnwidth]{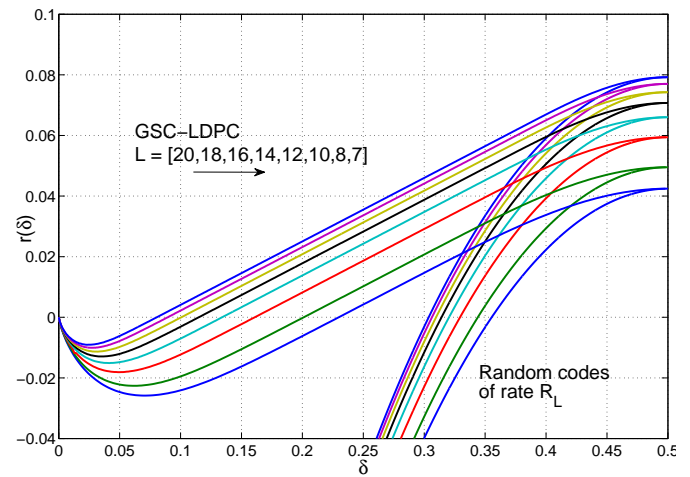}
\end{center}
\caption{Minimum distance growth rates of GSC-LDPC code ensembles and random linear codes of the corresponding rate.}\label{fig:termgrowth}
\end{figure}

Fig. \ref{fig:termgrowth} shows the asymptotic spectral shape functions for the GSC-LDPC code ensembles with termination factors $L=7,8,10,12,14,16,18,$ and $20$. Also shown are the asymptotic spectral shape functions for ``random'' codes of corresponding rate $R_L$ calculated using (see \cite{gal63})
\begin{equation}
r(\delta) = H(\delta)- (1-R_L)\ln(2),
\end{equation}
where $H(\delta)=-(1-\delta)\ln(1-\delta) - \delta\ln(\delta)$. We observe that the GSC-LDPC code ensembles are asymptotically good and have large minimum distance growth rates. This indicates that a long code based on this protograph has, with probability near one, a large minimum distance.  As $L$ increases, the design rate increases and the minimum distance growth rate decreases. This behavior is the same as was observed in the SPC case \cite{lmfc10}. 

%%%%%%%%%%%%%%%%%%%
%%%%%%%%%%%%%%%%%%%
\section{Free distance analysis of protograph-based GLDPCC code ensembles}\label{sec:freedist}
In Fig. \ref{fig:termgrowth} we saw that the minimum distance growth rates of GSC-LDPC codes decrease as the termination factor $L$ increases. However, since GSC-LDPC codes are terminated GLDPCC codes, a more appropriate distance measure for assessing the ML decoding performance of such codes is the \emph{free distance} growth rate of the GLDPCC ensemble. In this section, we first calculate the minimum distance growth rates for GTB-LDPC code ensembles and show that for sufficiently large termination factors, the growth rates coincide with those calculated for the GSC-LDPC code ensembles in Section \ref{sec:dist}. We then show that the growth rates of the GTB-LDPC code ensembles and GSC-LDPC code ensembles can be used to obtain lower and upper bounds on the free distance growth rate of the GLDPCC code ensemble, respectively.

\subsection{Minimum distance analysis of GTB-LDPC code ensembles}
We now consider terminating the protograph in Fig. \ref{fig:scprot} as a GTB-LDPC code with termination factor $\lambda$. Unlike the previous termination technique, this results in a $(2,7)$-regular protograph with design rate $R_\lambda = 1/7$ for all $\lambda$. The minimum distance growth rates of the GTB-LDPC code ensembles are presented in Fig. \ref{fig:termtbgrowth} alongside those corresponding to the GSC-LDPC code ensembles. We observe that the growth rates remain constant at $\delta_{min}=0.186$ (the growth rate of the original GLDPC block code ensemble) for $\lambda=1,2,\ldots,8,$ and then begin to decay to zero as $\lambda \rightarrow \infty$. Also, as a result of the convolutional structure, we observe that the GTB-LDPC and GSC-LDPC growth rates coincide for $L,\lambda\geq 10$. This is the same behavior that we observed for TB-LDPC and SC-LDPC codes with SPC constraints \cite{mpc13}.

\subsection{Free distance bounds for GLDPCC code ensembles}
Consider an ensemble of periodically time-varying GLDPCC codes with rate $R=1-b_cM_\mathcal{C}/b_v$ and period $T$ constructed from a convolutional protograph with base matrix $\binf$ (see (\ref{convbase})) as described in Section \ref{sec:convproto}.  Using a modification of the proof techniques in \cite{mpc13,tzc10}, it is possible to show that the average free distance of this ensemble can be bounded below by the average minimum distance of an ensemble of GTB-LDPC codes derived from the base matrix $\btb$ (see (\ref{basetb})) with termination factor $\lambda=T$ . Here, we show that the average free distance of the GLDPCC ensemble can also be bounded above by the average minimum distance of the ensemble of GSC-LDPC codes derived from the base matrix $\bterm$ (see (\ref{termbase})) with termination factor $L=T$. %A similar technique was used in \cite{stl+07}.% using the so-called `segment distance' to obtain a lower bound on the free distance of an ensemble of $(J,K)$-regular LDPCC codes. 

\newtheorem{freedist}{Theorem}
\begin{freedist}
Consider a rate $R=1-b_cM_\mathcal{C}/b_v$ unterminated, periodically
time-varying GLDPCC code ensemble with memory $m_s$, decoding constraint length
$\nu_{s}=N(m_s+1)b_v$, and period $T$ derived from $\binf$. Let $\overline{d}^{(L)}_{min}$ be
the average minimum distance of the GSC-LDPC
code ensemble with block length $n=L N b_v$ and termination factor
$L$. Then the ensemble average free distance $\overline{d}_{free}^{(T)}$ of the unterminated
convolutional code ensemble is bounded above by $\overline{d}_{min}^{(L)}$ for termination
factor $L=T$, i.e.,\vspace{-1mm}
\begin{equation}\label{dfreebound}
    \overline{d}_{free}^{(T)} \leq \overline{d}^{(T)}_{min}.\vspace{1mm}
\end{equation}
\end{freedist}
\emph{Sketch of proof}. There is a one-to-one relationship between members of the periodically time-varying GLDPCC code ensemble and members of the corresponding GSC-LDPC code ensemble with termination factor $L=T$. For any such pair of codes, every codeword $\mathbf{x}=[\begin{array}{cccc} x_0 & x_1 & \cdots & x_{L N b_v-1}\end{array}]$ in the GSC-LDPC (terminated convolutional) code can also be viewed as a codeword $\mathbf{x}_{[0,\infty]}=[\begin{array}{cccccc} x_0 & x_1 & \cdots & x_{L N b_v-1}&0&\cdots\end{array}]$ in the unterminated code. It follows that the free distance $d_{free}^{(T)}$ of the unterminated code cannot be larger than the minimum distance $d_{min}^{(T)}$ of the terminated code. %Further, a non-zero codeword sequence of the convolutional code could exist with Hamming weight less than $d_{min}^{(T)}$. 
The ensemble average result $\overline{d}_{free}^{(T)} \leq \overline{d}^{(T)}_{min}$ then follows directly. \hfill $\Box$

Since there is no danger of ambiguity, we will henceforth drop the overline notation when discussing ensemble average distance measures. 

\subsection{Free distance growth rates of GLDPCC code ensembles}\label{sec:growth}

For GLDPCC codes, %conventionally defined as the null space of a sparse parity-check matrix $\mathbf{H}_{[0,\infty]}$, 
it is natural to define the free distance growth rate with respect to the decoding constraint length $\nu_s$, i.e., as the ratio of the free distance $d_{free}$ to $\nu_s$.  

By bounding $d_{free}^{(T)}$ using (\ref{dfreebound}), we obtain an upper bound on the free distance growth rate as\vspace{-2mm}
\begin{equation}\label{ub}
\delta_{free}^{(T)}=\frac{d_{free}^{(T)}}{\nu_s} \leq \frac{\hat{\delta}_{min}^{(T)}T}{(m_s+1)},\vspace{-2mm}
\end{equation}
where $\hat{\delta}_{min}^{(T)} = {d_{min}^{(T)}}/{n}={d_{min}^{(T)}}/{(NT  b_v)}$ is the minimum distance growth rate of GSC-LDPC code ensemble with termination factor $L=T$ and base matrix $\mathbf{B}_{[0,T-1]}$. Similarly, using a similar argument to that presented in \cite{mpc13}, we have
\begin{equation}\label{lb}
\delta_{free}^{(T)}\geq \frac{\check{\delta}_{min}^{(T)}T}{(m_s+1)},\vspace{-2.5mm}
\end{equation}
where $\check{\delta}_{min}^{(T)}$ is the minimum distance growth rate of the GTB-LDPC code ensemble with tail-biting termination factor $\lambda=T$ and base matrix $\btb$.

The free distance growth rate $\delta_{free}^{(T)}$ that we bound from above using (\ref{ub}) is, by definition, an existence-type lower bound on the free distance of most members of the ensemble, i.e., with high probability a randomly chosen code from the ensemble has minimum free distance at least as large as ${\delta}_{free}^{(T)}\nu_s$ as $\nu_s\rightarrow \infty$. Note that the free distance growth rate may also be calculated with respect to the encoding constraint length $\nu_e$, which corresponds to the maximum number of transmitted symbols that can be affected by a single nonzero block of information digits. As a result of normalizing by the decoding constraint length, it is possible to have free distance growth rates larger than $0.5$. For further details, see \cite{mpc13}.

\subsection{Numerical results}
\begin{figure}[h]
\begin{center}
\includegraphics[width=\columnwidth]{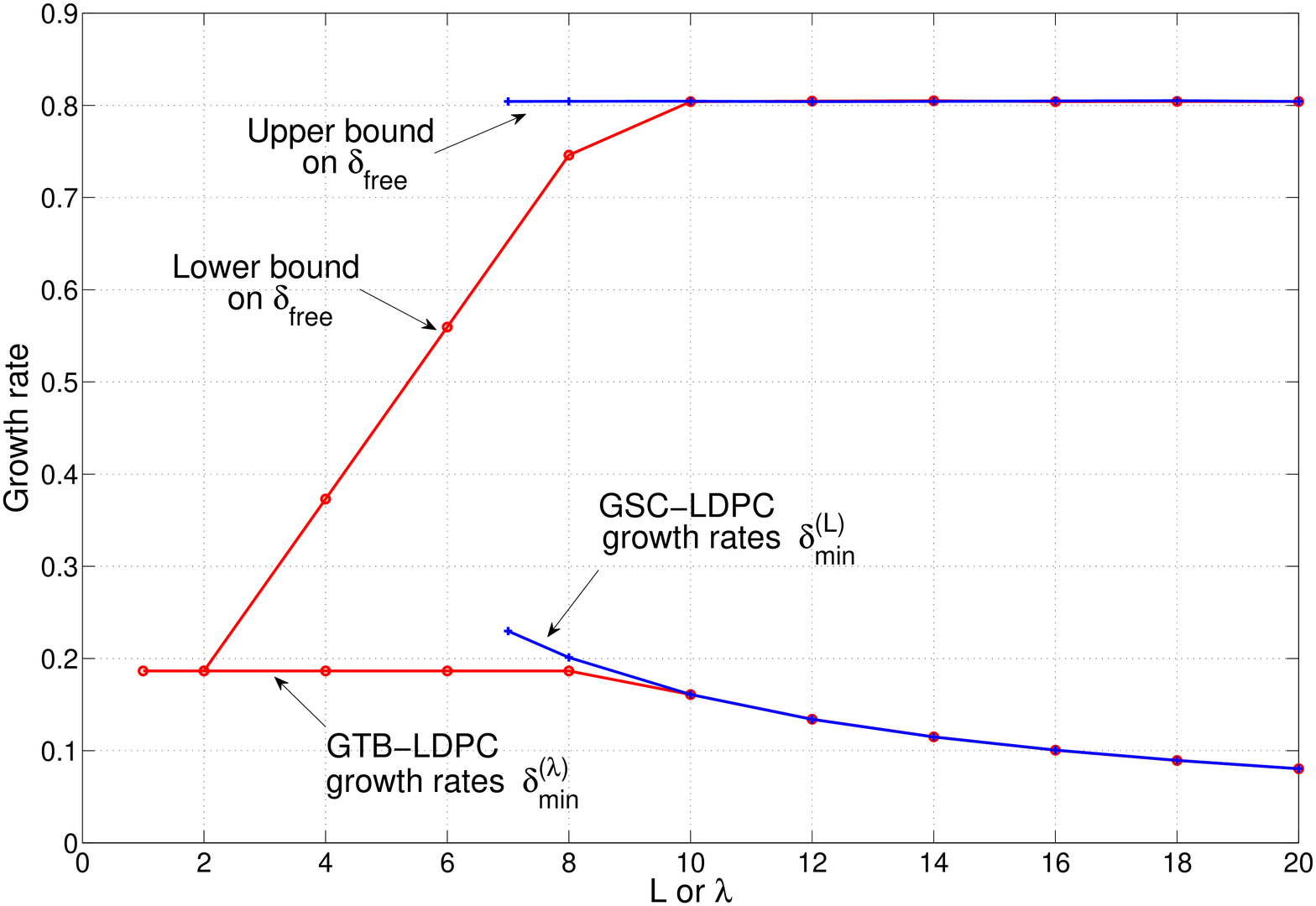}
\end{center}
\caption{Minimum distance growth rates of GSC-LDPC code ensembles and GTB-LDPC code ensembles and calculated upper and lower bounds on the free distance growth rates of the associated periodically time-varying GLDPCC code ensembles.}\label{fig:termtbgrowth}
\end{figure}

As an example, we consider once more the $(2,7)$-regular GLDPCC code ensemble with memory $m_s=1$ and rate $R=1/7$ depicted in Fig. \ref{fig:scprot}. For this case, we calculate the upper bound on the free distance growth rate of the periodically time-varying GLDPCC code ensemble as $\delta_{free}^{(T)} \leq \hat{\delta}^{(T)}_{min}T/2$ using (\ref{ub}) for termination factors $L=T\geq 7$. Fig. \ref{fig:termtbgrowth} displays the minimum distance growth rates $\hat{\delta}^{(L)}_{min}$ of the GSC-LDPC code ensembles defined by $\mathbf{B}_{[0,L-1]}$ for $L=7,8,10,12,\ldots,20$ that were calculated using the technique proposed in \cite{adr11} and the associated upper bounds on the GLDPCC code ensemble growth rates $\delta_{free}^{(T)} \leq \hat{\delta}^{(T)}_{min}T/2$ for $L=T$. Also shown are the minimum distance growth rates $\check{\delta}^{(\lambda)}_{min}$ of the GTB-LDPC code ensembles defined by base matrix $\mathbf{B}^{(\lambda)}_{tb}$ for $\lambda=1,2,4,\ldots,20$ and the associated lower bounds on the GLDPCC code ensemble growth rates $\delta_{free}^{(T)} \geq \check{\delta}^{(T)}_{min}T/2$ for $\lambda=T$ calculated using (\ref{lb}).

We observe that the calculated GTB-LDPC code ensemble minimum distance growth rates $\check{\delta}_{min}^{(\lambda)}$ remain constant for $\lambda=1, \ldots, 8$ and then start to decrease as the termination factor $\lambda$ grows, tending to zero as $\lambda$ tends to infinity. Correspondingly, as $\lambda$ exceeds $8$, the lower bound calculated for $\delta_{free}^{(T)}$ levels off at $\delta_{free}^{(T)} \geq 0.805$. The calculated GSC-LDPC code ensemble minimum distance growth rates $\hat{\delta}_{min}^{(L)}$ are larger for small values of $L$ (where the rate loss is larger) and decrease monotonically to zero as $L\rightarrow \infty$. Using (\ref{ub}) to obtain an upper bound on $\delta_{free}^{(T)}$ we observe that, for $T\geq 10$, the upper and lower bounds coincide, indicating that, for these values of the period $T$, $\delta_{free}^{(T)}=0.805$, significantly larger than the underlying GLDPC block code minimum distance growth rate $\delta_{min}=0.186$. In addition, we note that, at the point where the upper and lower bounds on $\delta_{free}^{(T)}$ coincide, the minimum distance growth rates for both termination methods also coincide. Recall that the GTB-LDPC code ensembles all have rate $1/7$, wheras the rate of the GSC-LDPC code ensembles is a function of the termination factor $L$ given by (\ref{ratesc}). This general technique can be used to bound the free distance growth rate above and below for any regular or irregular periodically time-varying protograph-based GLDPCC code ensemble.

While large free distance growth rates are indicative of good ML decoding performance, when predicting the iterative decoding performance of a code ensemble in the high SNR region other graphical objects such as trapping sets, pseudocodewords, absorbing sets, etc., come into effect. %The evaluation of trapping set enumerators for GSC-LDPC codes is the subject of ongoing work, however, 
Based on results from the SPC case \cite{mpc13}, we would expect GSC-LDPC codes with large minimum/free distance growth rates to also have large trapping set growth rates, indicating good iterative decoding performance in the high SNR region. 

%%%%%%%%%%%%%%%%%%%%%%%%
%%%%%%%%%%%%%%%%%%%%%%%%

\section{Conclusions}
GSC-LDPC codes constructed from a protograph are known to have better iterative decoding thresholds than their block code counterparts, and, for large termination lengths, their thresholds coincide with the MAP decoding threshold of the underlying GLDPC block code ensemble. In this paper, we used an asymptotic weight enumerator analysis to show that GSC-LDPC code ensembles are also asymptotically good. We saw, using a $(2,7)$-regular GLDPC code as an example, that the corresponding GSC-LDPC code ensembles have large minimum distance growth rates for all computed values of $L$. This indicates that long codes chosen from these ensembles have, with probability near one, large minimum distances as well as excellent iterative decoding thresholds. Finally, we obtained asymptotic minimum distance growth rates for GTB-LDPC code ensembles and showed that the growth rates of GTB-LDPC and GSC-LDPC code ensembles can be used to obtain lower and upper bounds, respectively, on the free distance growth rate of the associated periodically time-varying GLDPCC code ensemble.
\section*{Acknowledgment}
This work was partially supported by NSF Grant CCF-1161754.



\begin{thebibliography}{10}
\providecommand{\url}[1]{#1}
\csname url@samestyle\endcsname
\providecommand{\newblock}{\relax}
\providecommand{\bibinfo}[2]{#2}
\providecommand{\BIBentrySTDinterwordspacing}{\spaceskip=0pt\relax}
\providecommand{\BIBentryALTinterwordstretchfactor}{4}
\providecommand{\BIBentryALTinterwordspacing}{\spaceskip=\fontdimen2\font plus
\BIBentryALTinterwordstretchfactor\fontdimen3\font minus
  \fontdimen4\font\relax}
\providecommand{\BIBforeignlanguage}[2]{{%
\expandafter\ifx\csname l@#1\endcsname\relax
\typeout{** WARNING: IEEEtran.bst: No hyphenation pattern has been}%
\typeout{** loaded for the language `#1'. Using the pattern for}%
\typeout{** the default language instead.}%
\else
\language=\csname l@#1\endcsname
\fi
#2}}
\providecommand{\BIBdecl}{\relax}
\BIBdecl

\bibitem{fz99}
A.~{Jim\'{e}nez Felstr\"{o}m} and {K. Sh. Zigangirov}, ``Time-varying periodic
  convolutional codes with low-density parity-check matrices,'' \emph{IEEE
  Trans. Inf. Theory}, vol.~45, no.~6, pp. 2181--2191, Sept.
  1999.

\bibitem{psvc11}
A.~E. Pusane, R.~Smarandache, P.~O. Vontobel, and D.~J. {Costello, Jr.},
  ``Deriving good {LDPC} convolutional codes from {LDPC} block codes,''
  \emph{IEEE Trans. Inf. Theory}, vol.~57, no.~2, pp. 835--857,
  Feb. 2011.

\bibitem{lscz10}
M.~Lentmaier, A.~Sridharan, D.~J. {Costello, Jr.}, and {K. Sh. Zigangirov},
  ``Iterative decoding threshold analysis for {LDPC} convolutional codes,''
  \emph{IEEE Trans. Inf.n Theory}, vol.~56, no.~10, pp.
  5274--5289, Oct. 2010.

\bibitem{lfzc09}
M.~Lentmaier, G.~P. Fettweis, {K. Sh. Zigangirov}, and D.~J. {Costello, Jr.},
  ``Approaching capacity with asymptotically regular {LDPC} codes,'' in
  \emph{Proc. Inf. Theory and App. Workshop}, San Diego, CA,
  Feb. 2009.

\bibitem{kru11}
S.~Kudekar, T.~J. Richardson, and R.~L. Urbanke, ``Threshold saturation via
  spatial coupling: why convolutional {LDPC} ensembles perform so well over the
  {BEC},'' \emph{IEEE Trans. Inf. Theory}, vol.~57, no.~2, pp.
  803--834, Feb. 2011.

\bibitem{kru12}
\BIBentryALTinterwordspacing
S.~Kudekar, T.~Richardson, and R.~Urbanke, ``Spatially coupled ensembles
  universally achieve capacity under belief propagation,'' 2012. [Online].
  Available: \url{http://arxiv.org/abs/1201.2999}
\BIBentrySTDinterwordspacing

\bibitem{stl+07}
A.~Sridharan, D.~Truhachev, M.~Lentmaier, D.~J. {Costello, Jr.}, and {K. Sh.
  Zigangirov}, ``Distance bounds for an ensemble of {LDPC} convolutional
  codes,'' \emph{IEEE Trans. Inf. Theory}, vol.~53, no.~12, pp.
  4537--4555, Dec. 2007.

\bibitem{mpc13}
D.~G.~M. Mitchell, A.~E. Pusane, and D.~J. {Costello, Jr.}, ``Minimum distance
  and trapping set analysis of protograph-based {LDPC} convolutional codes,''
  \emph{IEEE Trans. Inf. Theory}, vol.~59, no.~1, pp. 254--281,
  Jan. 2013.

\bibitem{tan81}
R.~M. Tanner, ``A recursive approach to low complexity codes,'' \emph{IEEE
  Trans. Inf. Theory}, vol.~27, no.~5, pp. 533--547, Sept.
  1981.

\bibitem{lz99}
M.~Lentmaier and {K. Sh. Zigangirov}, ``On generalized low-density parity-check
  codes based on {Hamming} component codes,'' \emph{IEEE Comm.
  Letters}, vol.~8, no.~8, pp. 248--250, Aug. 1999.

\bibitem{bpz99}
J.~J. Boutros, O.~Pothier, and G.~Z\'{e}mor, ``Generalized low density {Tanner}
  codes,'' in \emph{Proc. IEEE Int. Conf. Comm.},
  Vancouver, Canada, June 1999.

\bibitem{lrc08}
G.~Liva, W.~E. Ryan, and M.~Chiani, ``Quasi-cyclic generalized {LDPC} codes
  with low error floors,'' \emph{IEEE Trans. Comm.}, vol.~56,
  no.~1, pp. 49--57, Jan. 2008.

\bibitem{lf10}
M.~Lentmaier and G.~Fettweis, ``On the thresholds of generalized {LDPC}
  convolutional codes based on protographs,'' in \emph{Proc. IEEE International
  Symposium on Information Theory}, Austin, TX, July 2010.

\bibitem{adr11}
S.~{Abu-Surra}, D.~Divsalar, and W.~E. Ryan, ``Enumerators for protograph-based
  ensembles of {LDPC} and generalized {LDPC} codes,'' \emph{IEEE Trans. Inf. Theory}, vol.~57, no.~2, pp. 858--886, Feb. 2011.

\bibitem{tzc10}
D.~Truhachev, {K. Sh. Zigangirov}, and D.~J. {Costello, Jr.}, ``Distance bounds
  for periodically time-varying and tail-biting {LDPC} convolutional codes,''
  \emph{IEEE Trans. Inf. Theory}, vol.~56, no.~9, pp.
  4301--4308, 2010.

\bibitem{tho03}
J.~Thorpe, ``Low-density parity-check ({LDPC}) codes constructed from
  protographs,'' Jet Propulsion Laboratory, Pasadena, CA, INP Progress Report
  42-154, Aug. 2003.

\bibitem{st79}
G.~Solomon and H.~C.~A. Tilborg, ``A connection between block and convolutional
  codes,'' \emph{SIAM Journal on App. Math.}, vol.~37, no.~2, pp.
  358--369, Oct. 1979.

\bibitem{mw86}
H.~H. Ma and J.~K. Wolf, ``On tail biting convolutional codes,'' \emph{IEEE
  Trans. Comm.}, vol.~34, no.~2, pp. 104--111, Feb. 1986.

\bibitem{gal63}
R.~G. Gallager, ``Low-density parity-check codes,'' Ph.D. dissertation,
  Massachusetts Institute of Technology, Cambridge, MA, 1963.

\bibitem{lmfc10}
M.~Lentmaier, D.~G.~M. Mitchell, G.~P. Fettweis, and D.~J. {Costello, Jr.},
  ``Asymptotically regular {LDPC} codes with linear distance growth and
  thresholds close to capacity,'' in \emph{Proc. Inf. Theory and
  App. Workshop}, San Diego, CA, Feb. 2010.

\end{thebibliography}
\end{document}